\documentclass[preprint,nofootinbib,amsmath,amssymb,aps]{revtex4-2}
\usepackage{fancyhdr}

\usepackage{setspace}
\usepackage[hang]{footmisc}
\usepackage{lipsum}
\begin{document}

\title{Angular reduction in multiparticle matrix elements}
\thanks{Originally published {\bf J Math Phys 30}, 2797 (1989)}
\thanks{This edition has typographic corrections, some rephrasings and a few added comments.}
\author{D.R. Lehman \& W.C. Parke}
\affiliation{The George Washington University, Washington, DC 20052}
\begin{abstract}
A general method for the reduction of coupled spherical harmonic products is
presented. When the total angular coupling is zero, the reduction leads to
an explicitly real expression in the scalar products within the unit vector
arguments of the spherical harmonics. For non-scalar couplings, the reduction
gives Cartesian tensor forms for the spherical harmonic products, with tensors
built from the physical vectors in the original expression. The reduction
for arbitrary couplings is given in closed form, making it amenable to
symbolic manipulation on a computer. The final expressions do not depend on
a special choice of coordinate axes, nor do they contain azimuthal quantum
number summations, nor do they have complex tensor terms for couplings to a
scalar. Consequently, they are easily interpretable from the properties of
the physical vectors they contain.
\end{abstract}

\maketitle

\newpage 

\fancyhf{}

\renewcommand{\footrulewidth}{0pt} \renewcommand{\headrulewidth}{0pt}


\rfoot{{\tiny }}

\affiliation{Physics Department, The George Washington University,
Washington, DC 20052}

\section{Introduction}

A common occurrence in quantum mechanical calculations for multiparticle
systems is the product of several spherical harmonics coming from the operators
and eigenstates of particle or cluster wave functions. For example, in
three-body models of $^{6}$Li, the quadrupole form factor begins with up to
five spherical harmonics coupled to zero total angular momentum, each with a
different argument (two each in the initial and final states, one in the
quadrupole operator). There are a variety of ways to evaluate transition
amplitudes and expectation values involving these products. This paper will
present an alternative that can be applied to arbitrary tensor couplings. When
those tensors are built from physical vectors in the problem, the method leads
to scalar couplings expressed as polynomials of the scalar products of those
vectors.

Several methods for handling a series of spherical harmonic couplings have been
suggested in the literature. One technique applies when only three are coupled,
and makes use of the freedom of choice for the orientation of the spatial axis
system. One of the spherical harmonic argument vectors is aligned with the
azimuthal quantization $z$ axis, and another pair defines the $xz$ plane. (See,
for example, the paper by Balian and Brezin.\footnote{R. Balian and E. Brezin,
{\bf Nuovo Cimento B 61}, 403 (1969).}) Putting coplanar vectors all in the $xy$ plane
also simplifies the explicit form of the spherical harmonics. In either case, a
sum over azimuthal quantum numbers remains for scalar expressions. An extension
of the above method takes advantage of the three-dimensional character of the
underlying space. The vector argument within any spherical harmonic in a
product expression is written in terms of any three independent vectors in the
problem. Those spherical harmonics with an argument direction determined by a
pair of vectors can be expanded as a product of spherical harmonics in each of
these vectors.\footnote{W. Kohn and N. Rostoker, {\bf Phys. Rev. 94}, 1111 (1954); see
also M. Danos and L. C. Maximon, {\bf J. Math. Phys. 6}, 766 (1965) for further
references going back to Lord Rayleigh.} Spherical harmonics with the same
argument are then combined. The result of the reduction will be a sum over
products of no more than three spherical harmonics in three different solid
angles. The technique described above can then be applied to these remaining
spherical harmonics.

For the case of a pair of coupled spherical harmonics with angular arguments
determined by two different unit vectors $\widehat{{\mathbf a}}$ and
$\widehat{{\mathbf b}}$, each spherical harmonic with high angular indices
($l_{1}$, and $l_{2}$) coupled to a total angular momentum of low angular index
(such as $L = 0, 1, 2, $ or $ 3$), it is possible\footnote{This method has been used
by J. L. Friar and G. L. Payne in two- and three-body calculations; for
details, see J. L. Friar and G. L. Payne, {\bf Phys. Rev. C 38}, 1 (1988).
\label{FP}} to
express the coupled pair in terms of a basis set of pair-coupled spherical
harmonics each with minimal angular index, times Legendre functions of argument
$\widehat{{\mathbf a}}\cdot \widehat{\mathbf b}$. Such results will turn out to be
special cases of the method given in the following.

In this paper, we wish
to present a general method for the reduction of products of spherical
harmonics which we have been using for some years.\footnote{The method was
originally developed by one of the authors (DRL) in conjunction with the
derivation of the three-body, bound-state equations for $\, ^6$He and $\, ^6$Li
[A. Ghovanlou and D. R. Lehman, {\bf Phys. Rev. C 9}, 1730 (1973); D. R. Lehman, M.
Rai, and A. Ghovanlou, {\bf Phys. Rev. C 17}, 744 (1978) ]. For the work on the $A = 6$
system, the method was worked out for angular-momentum values up to $l = 5$, and
used by DRL and his collaborators in numerous applications since that time [for
example, D. R. Lehman and M. Rajan, {\bf Phys. Rev. C 25}, 2743 (1982); B. F. Gibson
and D. R. Lehman, {\bf Phys. Rev. C 29}, 1017 (1984) ]. Recently, in association with
our work on the $\, ^{6}$Li quadrupole form factor with A. Eskandarian [A.
Eskandarian, D. R. Lehman, and W. C. Parke, {\bf Phys. Rev. C 38}, 2341 (1988)],
where the method was used to obtain programmable expressions for five spherical
harmonics coupled to zero, WCP generalized the method to arbitrary $l$ and
derived the irreducible decomposition of a product of two irreducible Cartesian
tensors of any rank.} When the total angular coupling is zero, the reduction
leads to an explicitly real expression in the scalar dot products of the vector
arguments of the original spherical harmonics. For non-scalar couplings, the
reduction gives Cartesian tensor forms for the spherical harmonic products;
tensors built from the physical vectors in the original problem. The advantages
of the method are the following: (1) The result is readily interpretable from
the known properties of the physical vectors it contains. (2) No special choice
of coordinate axes are needed. (3) The final expression contains no azimuthal
quantum number summations and no complex terms for couplings to a scalar. (4)
The reduction for arbitrary couplings can be given in closed form, making it
easily programmable in a computer calculation. As there are no spherical
harmonic origin-shift expansions, numerical convergence problems associated
with this re-expansion are avoided.

Section II introduces how the reduction of
the scalar couplings of spherical harmonics can lead to simple results in terms
of the corresponding vector dot-product expression. In Sec. III, we set up a
method for transforming between Cartesian and spherical tensors. Section IV
gives the general results for expanding the coupling of Cartesian tensors into
an irreducible tensor sum. A by-product of this work is a general formula for
the Cartesian Clebsch-Gordan coefficients. Section V shows how the Cartesian
coupling can reduce arbitrarily coupled spherical harmonics with different
arguments, using a few simple rules. Finally, Sec. VI summarizes our results.

\section{Examples of Scalar Coupling Reductions}

As a way of introducing the general scheme for Cartesian recoupling,
consider the following expression:
\begin{equation}
\lbrack Y^{\left[ 2\right] }(\widehat{{\mathbf a}})\times \lbrack Y^{\left[
1\right] }\left( \widehat{{\mathbf c}}\right) \times Y^{\left[ 1\right]
}(\widehat{{\mathbf d}})]^{\left[ 2\right] }]^{\left[ 0\right] }.  \label{eq1}
\end{equation}
We use here the angular coupling notation of Fano and Racah,\footnote{U. Fano
and G. Racah, {\bf Irreducible Tensorial Sets} (Academic, New York, 1959), pp. 36-38.} i.e.,
\begin{equation}
\left[ A^{\left[ l_{1}\right] }\times B^{\left[ l_{2}\right] }\right]
_{m_{3}}^{\left[ l_{3}\right] }=\sum_{m_1,m_2}^{m_{1}+m_{2}=m_{3}}\left\langle
l_{1}m_{1}l_{2}m_{2}\left| l_{3}m_{3}\right. \right\rangle A_{m_{1}}^{\left[
l_{1}\right] }B_{m_{2}}^{\left[ l_{2}\right] }\,.  \label{eq2}
\end{equation}
The phases for the `contrastandard' $Y_{m}^{\left[ l\right] }$ spherical
harmonics are fixed by
\begin{equation}
Y_{m}^{\left[ l\right] }=\left( -i\right) ^{l}Y_{lm} \ , \label{eq3}
\end{equation}
which insures that the $Y_{m}^{\left[ l\right] }$ behave as the eigenstates
of $L^{2}$ and $L_{z}$ under conjugation and time reversal according to
\begin{equation}
\psi _{m}^{\left[ l\right] \ast }=\left( -1\right) ^{l+m}\psi _{-m}^{\left[
l\right] } \ . \label{eq4}
\end{equation}
As emphasized by Danos, this phase choice also has the advantage of eliminating explicit phase factors
in matrix element angular recoupling algebra.\footnote{M. Danos, {\bf Ann. Phys.
63}, 319 (1971); D. R. Lehman and J. S. O'Connell, ``Graphical Recoupling of
Angular Momenta,'' {\bf National Bureau of Standards Monograph 136} (1973), p. 12; M.
Danos, V. Gillet, and M.Cauvin, {\bf Methods in Relativistic Nuclear Physics}
(North-Holland, Amsterdam, 1984), p. 59.},\footnote{A. R. Edmonds, {\bf Angular
Momentum in Quantum Mechanics} (Princeton UP, Princeton, NJ, 1960), Chaps. 6 and
7.}

It is widely known that the spherical harmonics $Y_{lm}(\widehat{{\mathbf a}})$
can be expressed in terms of
the symmetric traceless rank $l$ tensors made from the unit vector $\widehat{{\mathbf a}}$.
For example, in our notation,
we have in the cases of $l=1$ and $l=2$:
\begin{equation}
Y_{0}^{\left[ 1\right] }\left( \widehat{{\mathbf a}}\right) =+\left( -i\right)
\frac{\widehat{1}}{\sqrt{4\pi }}a_{3} \ , \label{eq5}
\end{equation}
\[
Y_{\pm 1}^{\left[ 1\right] }\left( \widehat{{\mathbf a}}\right) =\mp \left(
-i\right) \frac{\widehat{1}}{\sqrt{4\pi }}\frac{1}{\sqrt{2}}\left( a_{1}\pm
ia_{2}\right) ,
\]
and
\[
Y_{0}^{\left[ 2\right] }\left( \widehat{{\mathbf a}}\right) =+\left( -i\right)
^{2}\frac{\widehat{2}}{\sqrt{4\pi }}a_{33} \ ,
\]
\begin{equation}
Y_{\pm 1}^{\left[ 2\right] }\left( \widehat{{\mathbf a}}\right) =\mp \left(
-i\right) ^{2}\frac{\widehat{2}}{\sqrt{4\pi }}\sqrt{\frac{2}{3}}\left(
a_{13}\pm ia_{23}\right) \ , \label{eq6}
\end{equation}
\[
Y_{\pm 2}^{\left[ 2\right] }\left( \widehat{{\mathbf a}}\right) =+\left(
-i\right) ^{2}\frac{\widehat{2}}{\sqrt{4\pi }}\sqrt{\frac{1}{6}}\left(
a_{11}-a_{22}\pm 2ia_{12}\right) \ ,
\]
where \[\widehat{l}\equiv\sqrt{2l+1}\ ,\] and we define the second-rank
symmetric and traceless tensor $a_{ij}$ as
\begin{equation}
a_{ij}\equiv \frac{3}{2}\left( a_{i}a_{j}-\frac{1}{3}\delta _{ij}\right) \ .
\label{eq7}
\end{equation}

Our irreducible Cartesian tensors of rank $n$, $a_{i_1i_2\cdots i_n}$, are constructed from direct
products of a unit vector, and are normalized to make contraction
with that vector give the corresponding next lower rank tensor, until one
reaches $\widehat{{\mathbf a}}\cdot\widehat{{\mathbf a}}$, giving
unity.

The
expressions Eqs.\,(\ref{eq5}) and (\ref{eq6}) for the spherical harmonics can be checked using
\begin{equation}
\widehat{{\mathbf a}}=\sin \theta \cos \phi \,\widehat{\mathbf e}_{x}+\sin \theta \sin
\phi \,\widehat{\mathbf e}_{y}+\cos \theta \,\widehat{\mathbf e}_{z}  \label{eq8}
\end{equation}
and then comparing to known forms for the spherical harmonics\footnote{A. R.
Edmonds, {\it op. cit.}, Eq.\,(2.5.29).} written in terms of the spherical angles
$(\theta ,\phi )$.

As another example, the angular components of $[Y^{\left[ 1\right] }
\left( \widehat{{\mathbf c}}\right) \times Y^{\left[ 1\right]
}(\widehat{{\mathbf d}})]^{\left[ 2\right] }$ can be expressed in terms of the
components of the irreducible Cartesian tensor
\begin{equation}
Q\left( c,d\right) _{ij}\equiv\frac{3}{4}\left(
c_{i}d_{j}+c_{j}d_{i}-\frac{2}{3}\left( c\cdot d\right) \delta _{ij}\right) \,.
\label{eq9}
\end{equation}
The identity
\begin{equation}
\lbrack Y^{\left[ 1\right] }\left( \widehat{{\mathbf c}}\right) \times
Y^{\left[ 1\right] } (\widehat{{\mathbf c}})]_{m}^{\left[ 2\right]
}=\frac{\widehat{1}\widehat{1}}{\sqrt{4\pi }}\left| \left(
\begin{array}{ccc}
1 & 1 & 2 \\
0 & 0 & 0
\end{array}
\right) \right| Y_{m}^{\left[ 2\right] }(\widehat{{\mathbf c}})  \label{eq10}
\end{equation}
is employed to determine the constant factor in the results
\[
\lbrack Y^{\left[ 1\right] }\left( \widehat{{\mathbf c}}\right) \times
Y^{\left[ 1\right] }(\widehat{{\mathbf d}})]_{0}^{\left[ 2\right] }=+\left(
-i\right) ^{2}\frac{\widehat{2}}{\sqrt{4\pi }}Q\left( c,d\right) _{33} \ ,
\]
\begin{equation}
\lbrack Y^{\left[ 1\right] }\left( \widehat{{\mathbf c}}\right) \times
Y^{\left[ 1\right] }(\widehat{{\mathbf d}})]_{\pm 1}^{\left[ 2\right] }=\mp
\left( -i\right) ^{2}\frac{\widehat{2}}{\sqrt{4\pi }}\sqrt{\frac{2}{3}}\left(
Q\left( c,d\right) _{13}\pm iQ\left( c,d\right) _{23}\right) \ ,  \label{eq11}
\end{equation}
\[
\lbrack Y^{\left[ 1\right] }\left( \widehat{{\mathbf c}}\right) \times
Y^{\left[ 1\right] }(\widehat{{\mathbf d}})]_{\pm 2}^{\left[ 2\right] }=+\left(
-i\right) ^{2}\frac{\widehat{2}}{\sqrt{4\pi }}\sqrt{\frac{1}{6}}\left( Q\left(
c,d\right) _{11}-Q\left( c,d\right) _{22}\pm 2iQ\left( c,d\right) _{12}\right) \ .
\]
Using the traceless nature of $a_{ij}$ and $Q\left( c,d\right) _{ij}$, we
have
\[
a_{11}Q\left( c,d\right) _{22}+a_{22}Q\left( c,d\right) _{11}=a_{33}Q\left(
c,d\right) _{33}-a_{11}Q\left( c,d\right) _{11}-a_{22}Q\left( c,d\right)
_{22}
\]
so one finds
\begin{equation}
\left[ Y^{\left[ 2\right] }\left( \widehat{{\mathbf a}}\right) \times \lbrack
Y^{\left[ 1\right] }\left( \widehat{{\mathbf c}}\right) \times Y^{\left[
1\right] }(\widehat{{\mathbf d}})]^{\left[ 2\right] }\right] ^{\left[ 0\right]
}=\frac{\widehat{2}}{4\pi }\frac{2}{3}\sqrt{\frac{2\cdot 3}{5\cdot 4\pi
}}\sum_{i,j}a_{ij}Q\left( c,d\right) _{ij} \ . \label{eq12}
\end{equation}
The last factor, $\sum_{i,j}a_{ij}Q\left( c,d\right) _{ij}$, is just
\begin{equation}
\frac{3}{2}\widehat{{\mathbf a}}\cdot Q\left( c,d\right) \cdot
\widehat{{\mathbf a}}=\left( \frac{3}{2}\right) ^{2}\left\{ \left( a\cdot
c\right) \left( a\cdot d\right) -\frac{1}{3}\left( c\cdot d\right) \right\} \,.
\label{eq13}
\end{equation}
Aligning vector directions to help find the connection between the spherical
harmonic recoupling and the corresponding contracted Cartesian tensor
products will not work if the couplings have odd parity, such as in the
expression
\[
\left[ \left[ Y^{\left[ 2\right] }\left( \widehat{{\mathbf a}}\right) \times
Y^{\left[ 2\right] }\left( \widehat{{\mathbf b}}\right) \right] ^{\left[
1\right] }\times \lbrack Y^{\left[ 2\right] }\left( \widehat{{\mathbf
c}}\right) \times Y^{\left[ 2\right] }(\widehat{{\mathbf d}})]^{\left[ 1\right]
}\right] ^{\left[ 0\right] }\,.
\]
Two of the couplings above produce an axial vector from the direct product
of two tensors of rank two. If we define the pseudo-vector
\begin{equation}
R\left( a,b\right) _{i}\equiv\left( \frac{4}{9}\right) \sum_{jkl}\epsilon
_{ijk}a_{jl}b_{kl}=a\cdot b\left( a\times b\right) _{i}  \label{eq14}
\end{equation}
($\epsilon _{ijk}$ is the completely antisymmetric tensor in three
dimensions with $\epsilon _{123}=1$), then
\begin{eqnarray}
&&\left[ Y^{\left[ 2\right] }\left( \widehat{{\mathbf a}}\right) \times
Y^{\left[ 2\right] }\left( \widehat{{\mathbf b}}\right) \right] _{m}^{\left[
1\right] }
\label{eq15} \\
&=&\frac{1}{\sqrt{4\pi }}\sqrt{\frac{3\cdot 5}{2}}\left( -i\right)
\sqrt{\frac{3}{4\pi }}\left\{
\begin{array}{c}
R\left( a,b\right) _{3} \\
\mp \frac{1}{\sqrt{2}}\left( R\left( a,b\right) _{1}\pm R\left( a,b\right)
_{2}\right)
\end{array}
\right.
\begin{array}{c}
\left( m=0\right)  \\
\left( m=\pm 1\right) \ .
\end{array}
\nonumber
\end{eqnarray}
In this odd parity case, the coefficient in the expression
can be determined by the explicit Clebsch-Gordan recoupling of the spherical
harmonics with total azimuthal quantum number $m=0$. We now write the double
pair coupling to zero as
\begin{eqnarray}
&&\left[ \left[ Y^{\left[ 2\right] }\left( \widehat{{\mathbf a}}\right) \times
Y^{\left[ 2\right] }\left( \widehat{{\mathbf b}}\right) \right] ^{\left[
1\right] }\times \lbrack Y^{\left[ 2\right] }\left( \widehat{{\mathbf
c}}\right) \times Y^{\left[ 2\right] }(\widehat{{\mathbf d}})]^{\left[ 1\right]
}\right] ^{\left[ 0\right] }
\label{eq16} \\
&=&\left(\frac{\sqrt{3}}{4\pi}\right)\left(\frac{1}{\sqrt{4\pi}}\sqrt{\frac{3\cdot 5}{2}}
\right)^2\sum_{i}R\left( a,b\right) _{i}R\left( a,b\right) _{i} \ . \nonumber
\end{eqnarray}
The summed factor above becomes
\[(\widehat{\mathbf a}\cdot \widehat{\mathbf b})(\widehat{\mathbf d}\cdot \widehat{\mathbf e})\left[(\widehat{\mathbf a}\times \widehat{\mathbf b})\cdot (\widehat{\mathbf c}\times
\widehat{\mathbf d})\right]=(\widehat{\mathbf a}\cdot \widehat{\mathbf b})(\widehat{\mathbf d}\cdot \widehat{\mathbf e})\left[(\widehat{\mathbf a}\cdot \widehat{\mathbf c})(\widehat{\mathbf b}\cdot \widehat{\mathbf d})-(\widehat{\mathbf a}\cdot \widehat{\mathbf d})(\widehat{\mathbf b}\cdot \widehat{\mathbf c})\right].\]

In Sec. V, we show that expressions such as Eqs.\,(\ref{eq12}) and (\ref{eq16}) can be
written by inspection for arbitrary couplings.

\section{General Transformation between Irreducible Spherical and Cartesian
Tensors}

In this section, we will find a covariant connection between spherical and
Cartesian tensor components of arbitrary rank. This will lead to the
generalization of the results of Sec. II to arbitrary couplings of spherical
harmonics. To establish our notation, we first review the connection between
the generators of rotations and angular momentum. An orthonormal basis
$\widehat{e}_{i}$, in Euclidean three-space can be defined through the
infinitesimal displacements in that space by
\begin{equation}
d{\mathbf r}=\sum_{i}dx_{i}\,\widehat{{\mathbf e}}_{i} \ . \label{eq17}
\end{equation}
In a coordinate transformed frame, they become
\begin{equation}
\widehat{{\mathbf e}}_{i}^{\prime }=\sum_{j}\frac{dx_{j}}{dx_{i}^{\prime
}}\,\widehat{{\mathbf e}}_{j}\,\,.  \label{eq18}
\end{equation}

The condition
\begin{equation}
\sum_{i,j}\frac{dx_{k}^{^{\prime }}}{dx_{i}}\frac{dx_{l}^{^{\prime
}}}{dx_{j}}\delta _{ij}=\delta _{kl}  \label{eq19}
\end{equation}
makes the transformation a rotation. For infinitesimal orthogonal
transformations, Eqs. (18) and (19) give
\begin{equation}
\widehat{{\mathbf e}}_{i}^{^{\prime }}=\sum_{j}\left( \delta _{ij}+\sum
\epsilon _{ijk}n_{k}\delta \theta \right) \widehat{{\mathbf e}}_{j} \ ,
\label{eq20}
\end{equation}
where $\widehat{n}$ is a unit vector along the axis of rotation in the
right-hand sense and $\delta \theta $ is the rotation angle. Taking
$\mathcal{R} $ to be an element of the rotation group, an infinitesimal
rotation can be represented by
\begin{equation}
\mathcal{R}=\left( \mathcal{I}+i\sum_{k}S_{k}n_{k}\delta \theta \right) \ .
\label{eq21}
\end{equation}
Comparing with Eq.\,(\ref{eq20}), we can read the generator for infinitesimal
rotations of the Cartesian basis vectors to be the standard result:
\begin{equation}
\left( S_{k}\right) _{ij}=-i\epsilon _{ijk} \ . \label{eq22}
\end{equation}
Apart from Planck's constant, these are a representation for the angular
momentum operators for a spin-one field in quantum theory. However, the $z$
component of the angular momentum operator is usually taken as diagonal with
elements being the possible measured values of this $S_{z}$. The matrix $S_{z}$
is diagonalized by the unitary matrix
\begin{equation}
\left( \mathcal{U}_{mi}\right) =-i\left[
\begin{array}{ccc}
1/\sqrt{2} & -i/\sqrt{2} & 0 \\
0 & 0 & 1 \\
-1/\sqrt{2} & -i/\sqrt{2} & 0
\end{array}
\right] \ , \label{eq23}
\end{equation}
giving
\begin{equation}
\mathcal{U\,}S_{z}\,\mathcal{U}^{-1}=\left[
\begin{array}{ccc}
1 & 0 & 0 \\
0 & 0 & 0 \\
0 & 0 & -1
\end{array}
\right] \ . \label{eq24}
\end{equation}

The vector basis set in this contrastandard spherical representation is that
given by Danos\footnote{M. Danos, {\bf Ann. Phys. 63}, 319 (1971); D. R. Lehman and
J. S. O'Connell, ``Graphical Recoupling of Angular Momenta,'' {\bf National Bureau of
Standards Monograph 136} (1973), p. 12; M. Danos, V. Gillet, and M.Cauvin,
{\bf Methods in Relativistic Nuclear Physics} (North-Holland, Amsterdam, 1984), p.
59.}:
\[
\widehat{{\mathbf e}}_{m}^{\left[ 1\right]
}=\sum_{j}\mathcal{U}_{mi}\widehat{{\mathbf e}}_{i} \ ,
\]
\begin{equation}
\widehat{{\mathbf e}}_{\pm }^{\left[ 1\right] }=\pm \frac{i}{\sqrt{2}}\left(
\widehat{e}_{x}\pm \widehat{{\mathbf e}}_{y}\right)
\,,\,\,\,\,\,\widehat{{\mathbf e}}_{0}^{\left[ 1\right] }=-i\widehat{{\mathbf
e}}_{z}\,\,. \label{eq25}
\end{equation}
Furthermore,
\begin{equation}
\widehat{{\mathbf e}}_{m}^{\left[ 1\right] \,\dagger }\widehat{{\mathbf
e}}_{n}^{\left[ 1\right] \,}=\delta _{mn}\,\,,\,\,\,\,\,\,\,\sum_{m}\widehat{{\mathbf
e}}_{m}^{\left[ 1\right] }\widehat{{\mathbf e}}_{m}^{\left[ 1\right] \dagger
}={\mathbf 1}\,,  \label{eq26}
\end{equation}
where ${\mathbf 1}$ is the unit dyadic operator. (Note that these basis vectors differ from those of Fano and Racah\footnote{U. Fano and G. Racah,
{\it op. cit.}, p. 21.}. The Danos choice satisfies the conditions of Eq.\,(\ref{eq4}),
thus avoiding explicit phases when recoupling involves the angular unit vectors.)

The arbitrary phase in the unitary transformation has been taken to make the
spherical basis vectors conform with the conjugation property of angular
momentum eigenstates given in Eq.\,(\ref{eq4}). A contrastandard spherical tensor
carries a superscripted square bracket enclosing its rank index. Higher-weight
spherical tensors irreducible under the rotation group can be constructed from
angular couplings of the vector basis set:
\begin{equation}
\widehat{{\mathbf e}}_{m}^{\left[ l\right] }=\left[ \widehat{{\mathbf
e}}^{\left[ 1\right] }\times \widehat{{\mathbf e}}^{\left[ 1\right] }\times
\widehat{{\mathbf e}}^{\left[ 1\right] }\times \cdots \left( l\right) \cdots
\times \widehat{{\mathbf e}}^{\left[ 1\right] }\right] _{m}^{\left[ l\right]
}\,\,. \label{eq27}
\end{equation}
Individual pairwise couplings on the right-hand side of Eq.\,(\ref{eq27}) taken in
any order give the same result. This fact comes from the ``stretched'' form of
the tensor, i.e., it has the highest rank which can be constructed from $l $
vectors of rank 1. Explicitly, the Clebsch-Gordan products in Eq.\,(\ref{eq27})
give
\begin{eqnarray}
\widehat{{\mathbf e}}_{m}^{\left[ l\right] } &=&\left[ \frac{\left( l-m\right)
!\left(
l+m\right) !}{l!\left( 2l-1\right) !!}\right] ^{1/2} \times \nonumber \\
&&\sum_{m^{\prime }s\,from\,-1\,to\,1}\left[ \frac{1}{\left( 1-m_{1}\right)
!\left( 1+m_{1}\right) !\cdots \left( 1-m_{l}\right) !\left( 1+m_{l}\right)
!}\right] ^{1/2}\widehat{{\mathbf e}}_{m_{1}}^{\left[ 1\right] }\cdots
\widehat{{\mathbf e}}_{m_{l}}^{\left[ 1\right] }  \label{eq28}
\end{eqnarray}
The summation expression in Eq.\,(\ref{eq28}) implicitly depends on $m$, since the
coupled terms on the right-hand side of Eq.\,(\ref{eq27}) must have their
azimuthal quantum numbers add to $m$. The rank-$l$ tensors $\widehat{{\mathbf
e}}_{m}^{\left[ l\right] }$ satisfy
\begin{equation}
\widehat{{\mathbf e}}_{m}^{\left[ l\right] \,\dagger }\,\widehat{{\mathbf
e}}_{n}^{\left[ l\right] }=\delta _{mn}  \label{eq29}
\end{equation}
and
\begin{equation}
\sum_{m}\widehat{{\mathbf e}}_{m}^{\left[ l\right] }\,\widehat{{\mathbf
e}}_{m}^{\left[ l\right] \,\dagger }=\mathcal{P}^{\left[ l\right] } \ ,
\label{eq30}
\end{equation}
where $\mathcal{P}^{\left[ l\right] }$ is a projection operator on rank-$l$
Cartesian tensors which picks out only the irreducible part. The dot
products which appear between higher rank tensors imply contraction over all
Cartesian tensor indices.

A Cartesian tensor irreducible under the rotation group and of rank-$l$ must
be both completely symmetric in its $l$ indices and traceless. Suppose $%
T_{i_{1}\cdots i_{l}}$ is such a tensor. Then a natural connection between
this Cartesian tensor and its spherical representation is given by the
scalar expression:
\begin{eqnarray}
{\mathbf T} &=&\sum_{i^{\prime }s}T_{i_{1}i_{2}\cdots i_{l}}^{\left\{
l\right\}} \widehat{{\mathbf e}}_{i_{1}}\widehat{{\mathbf e}}_{i_{2}}\cdots
\widehat{{\mathbf e}}_{i_{l}}  \nonumber
\\
&=&\sum_{m}T_{m}^{\left[ l\right] }\widehat{{\mathbf e}}_{m}^{\left[ l\right]
\,\dagger }=\widehat{l}\left[ T^{\left[ l\right] }\times \widehat{{\mathbf
e}}^{\left[ l\right] }\right] ^{\left[ 0\right] }\,\,.  \label{eq31}
\end{eqnarray}

Contrastandard Cartesian tensors will be denoted by putting their rank index in
curly brackets. With Eq.\,(\ref{eq31}) the transformation coefficients between
Cartesian and spherical tensors become
\begin{equation}
\mathcal{U}_{mi_{1}\cdots i_{l}}^{\left[ l\right] }=\widehat{{\mathbf
e}}_{m}^{\left[ l\right] }\cdot \widehat{{\mathbf e}}_{i_{1}}\widehat{{\mathbf
e}}_{i_{2}}\cdots \widehat{{\mathbf e}}_{i_{l}}\,\,.  \label{eq32}
\end{equation}
Thus,
\begin{equation}
T_{i_{1}i_{2}\cdots i_{l}}^{\left\{ l\right\} }=\widehat{l}\left[ T^{\left[
l\right] }\times \mathcal{U}_{i_{1}\cdots i_{l}}^{\left[ l\right] }\right]
^{\left[ 0\right] }  \label{eq33}
\end{equation}
and
\begin{equation}
T_{m}^{\left[ l\right] }=\sum_{i^{\prime }s}T_{i_{1}i_{2}\cdots
i_{l}}^{\left\{ l\right\} }\mathcal{U}_{mi_{1}\cdots i_{l}}^{\left[ l\right]
}\,\,.  \label{eq34}
\end{equation}
The transformation coefficients satisfy the orthonormality conditions
\begin{equation}
\sum_{i^{\prime }s}\mathcal{U}_{mi_{1}\cdots i_{l}}^{\left[ l\right]
\,\,\dagger }\mathcal{U}_{ni_{1}\cdots i_{l}}^{\left[ l\right] }=\delta
_{mn}\,\,. \label{eq35}
\end{equation}
With
\begin{equation}
\mathcal{U}_{m3}^{\left[ 1\right] }=\left( -i\right) \delta _{m0} \ ,\label{eq36}
\end{equation}
we find from Eq.\,(\ref{eq28}),
\begin{equation}
\mathcal{U}_{m3\cdots 3}^{\left[ l\right] }=\left( -i\right) ^{l}\left[
\frac{l!}{\left( 2l-1\right) !!}\right] ^{1/2}\delta _{m0}\,\,. \label{eq37}
\end{equation}
The coefficients $\mathcal{U}_{mi_{1}\cdots i_{l}}^{\left[ l\right] }$, are
completely symmetric and traceless in the Cartesian indices $i_{1}$ to $i_{l}
$. Thus, they are irreducible in the space of both their spherical and
Cartesian indices.

We use Eq.\,(\ref{eq37}) to set the scale for normalization of Cartesian tensor
components relative to spherical ones:
\begin{equation}
T_{3\cdots 3}^{\left\{ l\right\} }=i^{l}\left[ \frac{l!}{\left( 2l-1\right)
!!}\right] ^{1/2}T_{0}^{\left[ l\right] }\,\,. \label{eq38}
\end{equation}

\section{Cartesian Tensor Recoupling}

A symmetric and traceless tensor of rank $l$ can be constructed from a unit
vector $\widehat{\mathbf{a}}$ in the form:
\begin{equation}
a^{\left\{ l\right\} }=\left[ \frac{\left( 2l-1\right) !!}{l!}\right]
\sum_{r=0}^{\left[ l/2\right] }\left( -1\right) ^{r}\left[ \frac{\left(
2l-2r-1\right) !!}{\left( 2l-1\right) !!}\right] \left\{ a\cdots \left(
l-2r\right) \cdots a\,\delta \cdots \left( r\right) \cdots \delta \right\} \,\,.
\label{eq39}
\end{equation}

These are the Cartesian equivalents of the spherical harmonics, which we will
refer to as `Cartesian harmonic tensors'. In this expression, as in Eq.\,(\ref{eq27}),
the parenthetical value between continuation dots shows the number
of repetitions of the factor shown before and after the dots. The $l$%
-Cartesian indices have been suppressed on $a^{\left\{ l\right\} }$ and in each
term of the summation. The $\delta $'s above are double-indexed Kronecker
deltas. The curly brackets in the right-hand side of Eq.\,(\ref{eq39}) direct that
the terms inside are to be summed over all permutations of the unsymmetrized
indices. For a given summation index $r$, there will be $\left[ l!/\left(
l-2r\right) !2^{r}r!\right] $\ such terms in the symmetrization bracket. Our
choice for normalization of $a^{\left\{ l\right\} }$ leads to (all vectors here,
even when not marked with a caret, are unit vectors)
\begin{equation}
\widehat{{\mathbf a}}^{\left\{ l\right\} }\cdot \widehat{{\mathbf b}}\cdots
\left( l-contractions\right) \cdots \widehat{{\mathbf b}}=P_{l}\left(
\widehat{{\mathbf a}}\cdot \widehat{{\mathbf b}}\right) \,, \label{eq40}
\end{equation}
where $\widehat{{\mathbf b}}$ is a second unit vector and $P_{l}\left(
\widehat{{\mathbf a}}\cdot \widehat{{\mathbf b}}\right) $ is the Legendre
polynomial. As an example, Eq.\,(\ref{eq39}) for $l=3$ becomes
\begin{equation}
a_{i_{1}i_{2}i_{3}}^{\left\{ 3\right\}
}=\frac{5}{2}a_{i_{1}}a_{i_{2}}a_{i_{3}}-\frac{1}{2}\left( a_{i_{1}}\delta
_{i_{2}i_{3}}+a_{i_{2}}\delta _{i_{3}i_{1}}+a_{i_{3}}\delta
_{i_{1}i_{2}}\right) \,\,. \label{eq41}
\end{equation}

By using Eq.\,(\ref{eq38}) and
\begin{equation}
Y_{0}^{\left[ l\right] }\left( \widehat{{\mathbf a}}\right) =\left( -i\right)
^{l} \frac{\widehat{l}}{\sqrt{4\pi }}P_{l}\left( \left( \widehat{{\mathbf
a}}\cdot \widehat{\mathbf{e}}_{3}\right) \right) \ , \label{eq42}
\end{equation}
it follows that the irreducible Cartesian tensors defined by Eq.\,(\ref{eq39}) are
related to the Cartesian transformed spherical harmonics by
\begin{equation}
Y^{\left\{ l\right\} }\left( \widehat{{\mathbf a}}\right)
=\frac{\widehat{l}}{\sqrt{4\pi }}\left[ \frac{l!}{\left( 2l-1\right) !!}\right]
^{1/2}a^{\left\{ l\right\} }\,\,. \label{eq43}
\end{equation}

Now consider the coupling of two irreducible tensors of rank $l_{1}$ and $%
l_{2}$. The result can be decomposed into a sum of irreducible tensors from
rank $\left| l_{1}-l_{2}\right| $ to $l_{1}+l_{2}$. This summation is well
known in the case of spherical tensors, giving a Clebsch-Gordan series. The
irreducible Cartesian tensors following from this decomposition must again
be completely symmetric and traceless. By explicitly constructing symmetric
and traceless tensors from the products of two irreducible tensors $%
A^{\left\{ l_{1}\right\} }$ and $B^{\left\{ l_{2}\right\} }$, it is
straightforward to show that the general form for decomposition of the irreducible rank $%
l_{3} $ Cartesian tensor is given by

\begin{equation}
\begin{array}{l}
\left[ A^{\left\{ l_{1}\right\} }\times B^{\left\{ l_{2}\right\} }\right]
^{\left\{ l_{3}\right\} } =C_{l_{1}l_{2}l_{3}}\left[ \frac{\left( \left(
l_{1}-l_{2}+l_{3}\right) /2\right) !\left( \left( l_{2}-l_{1}+l_{3}\right)
/2\right) !}{l_{3}!}\right] \times \\
 \\
 \sum_{r=0}^{\min \left[ l_{1}-k,l_{2}-k\right] }\left(
-1\right) ^{r}2^{r}\frac{\left( 2l_{3}-2r-1\right) !!}{\left( 2l_{3}-1\right)
!!}\left\{ A^{\left\{ l_{1}\right\} }\cdot \left( k+r\right)  B^{\left\{
l_{2}\right\} }\delta \cdots \left( r\right)\cdots \delta \right\}
\end{array} \label{eq44}
\end{equation}
when $l_{1}+l_{2}-l_{3}\equiv 2k$ is even, and by
\begin{equation}
\begin{array}{l}
\left[ A^{\left\{ l_{1}\right\} }\times B^{\left\{ l_{2}\right\} }\right]
^{\left\{ l_{3}\right\} } =D_{l_{1}l_{2}l_{3}}\left[ \frac{\left( \left(
l_{1}-l_{2}+l_{3}-1\right) /2\right) !\left( \left(
l_{2}-l_{1}+l_{3}-1\right) /2\right) !}{l_{3}!}\right] \frac{1}{\sqrt{2}}\,\times \\
 \\
\sum_{r=0}^{\min \left[ l_{1}-k^{\prime }-1,l_{2}-k^{\prime }-1\right]
 }\left( -1\right) ^{r}2^{r}\frac{\left( 2l_{3}-2r-1\right) !!}{\left( 2l_{3}-1\right)
 !!} \left\{ \epsilon :A^{\left\{ l_{1}\right\}
}\cdot \left( k^{\prime }+r\right)  B^{\left\{ l_{2}\right\} }\delta
\cdots \left( r\right) \cdots \delta \right\}
\end{array} \label{eq45}
\end{equation}
when $l_{1}+l_{2}-l_{3}\equiv 2k'+1$ is odd.

In these expressions, a dot on the left side of a parenthetical value $(k)$
and between two Cartesian tensors indicates a tensor contraction of order $k$.
A triple of dots on each side of a parenthetical value $(k)$ between
a tensor on each side indicates
one is to include a direct product of $k$ such tensors.
In addition, the colon indicates a double contraction of the form
\begin{equation}
\left( {\mathbf \epsilon :AB}\right) _{i_{1}i_{2}\cdots
i_{l}}=\sum_{j,k}\epsilon _{i_{1}jk}\ A_{ji_{2}\cdots }B_{k\cdots i_{l}}\,\,.
\label{eq46}
\end{equation}
In Eq.\,(\ref{eq44}), terms within the curly bracket are summed over permutations
of the indices across $A$, $B$, and $\delta $, leading to a symmetric tensor
with
\begin{equation}
\left[ \frac{l_{3}!}{\left( l_{1}-k-r\right) !\left( l_{2}-k-r\right)
!2^{r}r!}\right] \label{eq47}
\end{equation}
terms for each $r$, while for Eq.\,(\ref{eq45}), the symmetrization bracket gives
\begin{equation}
\left[ \frac{l_{3}!}{\left( l_{1}-k^{\prime }-r-1\right) !\left(
l_{2}-k^{\prime }-r-1\right) !2^{r}r!}\right] \label{eq48}
\end{equation}
terms for each $r$. The factors $C_{l_{1}l_{2}l_{3}}$ and $%
D_{l_{1}l_{2}l_{3}}$ will be determined by specializing the tensors in Eqs.
(44) and (45) to ones constructed from vectors. The square-bracketed
coefficients in Eqs. (44) and (45) are the inverse of the number of terms in
the first symmetrization bracket of the following summation. The $\left(
1/\sqrt{2}\right) $ factor in Eq.\,(\ref{eq45}) is inserted in anticipation of the
concurrence of the $C$ and $D$ coefficients.

For $A^{\left\{ l_{1}\right\} }=a^{\left\{ l_{1}\right\} }$ and $B^{\left\{
l_{2}\right\} }=a^{\left\{ l_{2}\right\} }$, the right-hand side of Eq.\,(\ref{eq44})
must be proportional to $a^{\left\{ l_{3}\right\} }$. With these
substitutions, the summations in Eq.\,(\ref{eq44}) can be performed, giving
\begin{equation}
\left[ a^{\left\{ l_{1}\right\} }\times a^{\left\{ l_{2}\right\} }\right]
^{\left\{ l_{3}\right\} }=C_{l_{1}l_{2}l_{3}}\frac{J_{1}!!J_{2}!!J_{3}!!\left(
J/2\right) !}{l_{3}!\left( 2l_{3}-1\right) !!}a^{\left\{ l_{3}\right\} }\,\,,
\label{eq49}
\end{equation}
where $J\equiv l_{1}+l_{2}+l_{3}$ and $J_{i}\equiv J-2l_{i}-1$, and $%
(-1)!!\equiv -1$.

With the spherical harmonic coupling identity
\begin{equation}
\left[ Y^{\left[ l_{1}\right] }\left( \widehat{{\mathbf c}}\right) \times
Y^{\left[ l_{2}\right] }\left( \widehat{{\mathbf c}}\right) \right]
_{m}^{\left[ l_{3}\right] }=\frac{\widehat{l}_{1}\widehat{l}_{2}}{\sqrt{4\pi
}}\left| \left(
\begin{array}{ccc}
l_{1} & l_{2} & l_{3} \\
0 & 0 & 0
\end{array}
\right) \right| Y_{m}^{\left[ l_{3}\right] }\left( \widehat{{\mathbf c}}\right)
\label{eq50}
\end{equation}
together with Eq.\,(\ref{eq43}), we find
\begin{equation}
C_{l_{1}l_{2}l_{3}}=\widehat{l_{3}}\left[ \frac{\left( 2l_{1}\right) \left(
2l_{2}\right) \left( 2l_{3}\right) }{\left( J_{1}+1\right) \left(
J_{2}+1\right) \left( J_{3}+1\right) \left( J+1\right) }\right] ^{1/2}
\label{eq51} \end{equation}

For odd $l_{1}+l_{2}+l_{3}$, one can compare relation (45) for $A^{\left\{
l_{1}\right\} }=a^{\left\{ l_{1}\right\} }$ and $B^{\left\{ l_{2}\right\}
}=b^{\left\{ l_{2}\right\} }$ with the corresponding spherical harmonic
coupling. Using the Clebsch-Gordan coefficients for $m_{3}=0$, it follows
(after some tedious algebra) that
\begin{equation}
D_{l_{1}l_{2}l_{3}}=C_{l_{1}l_{2}l_{3}}\,\,. \label{eq52}
\end{equation}

The relations (44) and (45) with (51) and (52) constitute an explicit
solution for the Clebsch-Gordan coefficients in an expansion of a product of
irreducible tensors in Cartesian form.

\section{Applications to Spherical Harmonic Couplings}

We now are in a position to reduce any set of spherical harmonic couplings to
Cartesian form. Repeated application of the pairwise coupling formula (44) and
(45) will necessarily lead to a Cartesian expression in the original vectors of
the problem. For couplings to a scalar, clearly the result will be a polynomial
in the scalar products of these vectors, with order no greater than the smaller
of the ranks of the two spherical harmonics entering with these vector
arguments. If the coupling is to a pseudo-scalar, a ``box'' product (e.g.,
${\mathbf a}\cdot \mathbf{b}\times \mathbf{c}$) of three independent vectors
must be an overall factor.

The reduction of an arbitrary series of spherical harmonic couplings proceeds
as follows: For each spherical harmonic, introduce the rescaling factors shown
in Eq.\,(\ref{eq43}). For each pair coupling, write the appropriate Cartesian
coupling as in Eq.\,(\ref{eq44}) or (\ref{eq45}). Finally, perform the indicated Cartesian
tensor contractions, starting with Eq.\,(\ref{eq39}) for each spherical harmonic.
In this process, the traceless nature of these tensors greatly simplifies the
reduction, since the Kronecker delta's within one such tensor contracted with
another irreducible tensor will vanish.

For example,\footnote{See Eq.\,(17b) of Friar and Payne, {\it op. cit.} We used
similar relations to derive the equations for the shell structure of the A = 6
ground state from three-body dynamics as given in the appendix of D. R. Lehman
and W. C. Parke, {\bf Phys. Rev. C 28}, 364 (1983).} the above method can be used to
show that
\begin{equation}
\left[ Y^{\left[ l\right] }\left( \widehat{{\mathbf a}}\right) \times Y^{\left[
l\right] }\left( \widehat{{\mathbf b}}\right) \right] _{m}^{\left[ 1\right]
}=\frac{\left( -i\right) }{4\pi }\left[ \frac{3\left( 2l+1\right) }{l\left(
l+1\right) }\right] ^{1/2}P_{l}^{\prime }\left[ \widehat{{\mathbf a}}\times
\widehat{{\mathbf b}}\right] _{m} \label{eq53a}
\end{equation}
and
\[
\left[ Y^{\left[ l-1\right] }\left( \widehat{{\mathbf a}}\right) \times
Y^{\left[ l\right] }\left( \widehat{{\mathbf b}}\right) \right] _{m}^{\left[
1\right] }=\frac{\left( -i\right) }{4\pi }\left[ \frac{3}{l}\right]
^{1/2}\left[ P_{l}^{\prime }\,\widehat{{\mathbf b}}_{m}-\left( \left( l-1\right)
P_{l-2}+\widehat{{\mathbf a}}\cdot \widehat{{\mathbf b}}\,P_{l-2}^{\prime
}\right) \widehat{{\mathbf a}}_{m}\right] \label{eq53b} \ ,
\]
where $P_{l}\left( \widehat{{\mathbf a}}\cdot \widehat{{\mathbf b}}\right) $ is
the Legendre function of order $l$ and $P_{l}^{\prime }$ is its derivative
with respect to its argument.
Similarly, higher-order couplings of the form $\left[ Y^{\left[ l_{1}\right]
}\left(\widehat{{\mathbf a}}\right) \times Y^{\left[ l_{2}\right] }\left(
\widehat{{\mathbf b}}\right) \right] ^{\left[ L\right] }$ for $L=2,3,...$ can
be expressed in terms of the order-$L$ ``stretched'' even or odd parity couplings
of the vectors $\widehat{{\mathbf a}}$ and $\widehat{{\mathbf b}}$ times
Legendre functions and their derivatives. They are most easily derived by
expanding the given form in terms of an independent set of stretched couplings
with unknown scalar coefficients, then contracting with each tensor of the set
to form scalar relations for the coefficients.

In matrix element calculations, spherical harmonic couplings to total
angular momentum of zero arise.
In these cases, we have found it convenient
to introduce a set of rules for generating the final scalar expression given
the initial coupling. These rules result from the Cartesian recoupling
formalism of the last section and are taken in a form which allows for an
easy verification of each step.

The rules are as follows:

Step (la): For each interior pair coupling of even parity, introduce the
Cartesian tensor factor
\begin{equation}
\begin{array}{l}
 Q_{l_{1}l_{2}l_{3}}\left( A,B\right) \equiv \left[
\frac{l_{1}!l_{2}!\left( 2l_{3}\right) !!\left( \left( J_{1}+1\right) /2\right)
!\left( \left( J_{2}+1\right) /2\right) !}{l_{3}!J_{1}!!J_{2}!!J_{3}!!\left(
J/2\right) !}\right] \times \\
 \\
\sum_{r=0}^{\min \left[ l_{1}-k,l_{2}-k\right] }\left( -1\right)
^{r}2^{r}\frac{\left( 2l_{3}-2r-1\right) !!}{\left( 2l_{3}-1\right) !!}\times
\left\{ A^{\left\{ l_{1}\right\} }\cdot \left( k+r\right) B^{\left\{
l_{2}\right\} }\delta \cdots \left( r\right) \cdots \delta \right\}
\end{array} \label{eq54}
\end{equation}
coming from the coupling in Eq.\,(\ref{eq44}). As before, $J\equiv l_{1}+l_{2}+l_{3},$
$J_{i}\equiv J-2l_{i}-1$, $2k=l_{1}+l_{2}-l_{3}$, $(-1)!!=1$, and the
bracketed terms contain an implicit symmetrization sum, with the number of
such terms given by the expression in (47). This rank-$l_{3}$, tensor has
been normalized so that when $A^{\left\{ l_{1}\right\} }=a^{\left\{
l_{1}\right\} }$ and $B^{\left\{ l_{2}\right\} }=b^{\left\{ l_{2}\right\} }$ \ ,
$Q$ reduces to $a^{\left\{ l_{3}\right\} }$. Thus the $Q$'s are a natural
generalization of the Cartesian harmonic tensors. Note also that
\[
Q_{l_{1}l_{2}l_{3}}\left( a,a\right) \cdot \left( l_{3}\right) {\mathbf
b}=P_{l_{3}}\left( \widehat{{\mathbf a}}\cdot \widehat{{\mathbf b}}\right)
\]
and so
\[
Q_{l_{1}l_{2}l_{3}}\left( a,a\right) \cdot \left( l_{3}\right) {\mathbf
a}=1\,\,.
\]

Step (1b): For each interior pair coupling of odd parity, introduce the
Cartesian tensor factor
\begin{equation}
\begin{array}{l}
R_{l_{1}l_{2}l_{3}}\left( A,B\right) \equiv \left[ \frac{2l_{1}!l_{2}!\left(
2l_{3}-1\right) !!\left( J_{1}/2\right) !\left( J_{2}/2\right) !}{\left(
l_{3}-1\right) !\left( J_{1}+1\right) !!\left( J_{2}+1\right) !!\left(
J+1\right) !!\left( \left( J+1\right) /2\right) !}\right] \\
 \\
\times \sum_{r=0}^{\min \left[ l_{1}-k^{\prime }-1,l_{2}-k^{\prime }-1\right] }
\left( -1\right) ^{r}2^{r}\frac{\left( 2l_{3}-2r-1\right) !!}{\left(
2l_{3}-1\right) !!}\left\{ \epsilon :A^{\left\{ l_{1}\right\} }\cdot \left(
k^{\prime }+r\right) B^{\left\{ l_{2}\right\} }\delta \cdots \left( r\right)
\cdots \delta \right\}
\end{array} \label{eq55}
\end{equation}
coming from the coupling in Eq.\,(\ref{eq45}) ($J\equiv l_{1}+l_{2}+l_{3}$, $%
J_{i}\equiv J-2l_{i}-1$, $2k^{\prime }+1=l_{1}+l_{2}-l_{3}$). As before, the
bracketed term contains an implicit symmetrization sum, with the number of
such terms given in (48). This tensor has been normalized so that, for $%
A^{\left\{ l_{1}\right\} }=a^{\left\{ l_{1}\right\} }$, $B^{\left\{
l_{2}\right\} }=b^{\left\{ l_{2}\right\} }$, and as the vector
$\widehat{{\mathbf b}}$ approaches $\widehat{{\mathbf a}}$, we have
\begin{equation}
\lim_{b\rightarrow a}\frac{\left| R_{l_{1}l_{2}l_{3}}\left( a,b\right) \cdot
\left( l_{3}-1\right) {\mathbf a}\right| }{\left| \widehat{{\mathbf a}}\times
\widehat{{\mathbf b}}\right| }=1 \label{eq56}
\end{equation}

Step (2a) : For even parity couplings, introduce a factor
\begin{eqnarray}
q_{l_{1}l_{2}l_{3}} &\equiv &\frac{\widehat{l}_{1}\widehat{l}_{2}}{\sqrt{4\pi
}}\left| \left(
\begin{array}{ccc}
l_{1} & l_{2} & l_{3} \\
0 & 0 & 0
\end{array}
\right) \right| \\
&=&\frac{\widehat{l}_{1}\widehat{l}_{2}}{\sqrt{4\pi }}\left[
\frac{J_{1}!!J_{2}!!J_{3}!!\left( J/2\right) !}{\left( \left( J_{1}+1\right)
/2\right) !\left( \left( J_{2}+1\right) /2\right) !\left( \left( J_{3}+1\right)
/2\right) !\left( J+1\right) !!}\right] ^{1/2}\,\,. \nonumber
\label{eq57}
\end{eqnarray}

Our normalization for $Q$ makes $q$ the same factor which one would
ordinarily use in coupling spherical harmonics with identical arguments.

Step (2b) : For odd parity couplings, introduce a factor
\begin{eqnarray}
r_{l_{1}l_{2}l_{3}} &\equiv &\frac{\widehat{l}_{1}\widehat{l}_{2}}{\sqrt{4\pi
}} \left[ \frac{l_{1}\left( l_{1}+1\right) l_{2}\left( l_{2}+1\right)
}{2l_{3}\cdot 2l_{3}}\right] ^{1/2}\left| \left(
\begin{array}{ccc}
l_{1} & l_{2} & l_{3} \nonumber \\
1 & -1 & 0
\end{array}
\right) \right| \\
&=&\frac{\widehat{l}_{1}\widehat{l}_{2}}{2l_{3}\sqrt{4\pi }}\left[ \frac{\left(
J_{1}+1\right) !!\left( J_{2}+1\right) !!\left( J_{3}+1\right) !!\left( \left(
J+1\right) /2\right) !}{\left( J_{1}/2\right) !\left( J_{2}/2\right) !\left(
J_{3}/2\right) !J!!}\right] ^{1/2} \label{eq58}
\end{eqnarray}

Step (3) : For the final $L\times L$ coupling to $0$, use a factor
\begin{equation}
S\equiv \frac{\widehat{L}}{4\pi }\frac{L!}{\left( 2L-1\right) !!} \label{eq59}
\end{equation}
and fully contract the final pair of Cartesian tensors. The factors in Eqs.
(12) and (16) have been arranged to exhibit these steps.

As another example, consider the fourfold coupling
\begin{equation}
\left[ \left[ Y^{\left[ 2\right] }\left( \widehat{{\mathbf a}}\right) Y^{\left[
2\right] }\left( \widehat{{\mathbf b}}\right) \right] ^{\left[ 2\right] }\times
\lbrack Y^{\left[ 1\right] }\left( \widehat{{\mathbf c}}\right) \times
Y^{\left[ 3\right] }(\widehat{{\mathbf d}})]^{\left[ 2\right] }\right] ^{\left[
0\right] }\,\,. \label{eq60}
\end{equation}
Performing each step on the above coupling (from right to left), we have
\begin{equation}
\frac{\sqrt{5}}{4\pi }\cdot\frac{2}{3}\cdot\sqrt{\frac{3\cdot 3}{5\cdot 4\pi
}}\sqrt{\frac{2\cdot 5}{7\cdot 4\pi }}Q_{222}\left( a,b\right) :Q_{132}\left(
c,d\right) , \label{eq61}
\end{equation}
where
\begin{equation}
Q_{222}\left( a,b\right) =\frac{9}{4}\left\{ \widehat{{\mathbf a}}\cdot
\widehat{{\mathbf b}}\left( \widehat{{\mathbf a}}\widehat{{\mathbf
b}}+\widehat{{\mathbf b}}\widehat{{\mathbf a}} -\frac{2}{3}\widehat{{\mathbf
a}}\cdot \widehat{{\mathbf b}}\delta \right) -\frac{2}{3}\left(
\widehat{{\mathbf a}}\widehat{{\mathbf a}}-\frac{1}{3}\delta \right)
-\frac{2}{3}\left( \widehat{{\mathbf b}}\widehat{{\mathbf b}}-\frac{1}{3}\delta
\right) \right\} \label{eq62}
\end{equation}
and
\begin{equation}
Q_{132}\left( c,d\right) =\frac{5}{2}\left\{ \widehat{{\mathbf c}}\cdot
\widehat{{\mathbf d}}\left( \widehat{{\mathbf d}}\widehat{{\mathbf
d}}-\frac{1}{3}\delta \right) +\frac{1}{5}\left( \widehat{{\mathbf
c}}\widehat{{\mathbf d}}+\widehat{{\mathbf d}}\widehat{{\mathbf
c}}-\frac{2}{3}\widehat{{\mathbf c}}\cdot \widehat{{\mathbf d}}\delta \right)
\right\} \,\,. \label{eq63}
\end{equation}
Making the last contraction to a scalar is simplified by noting that all
contractions of the Kronecker deltas in $Q_{222}\left( a,b\right) $ with
$Q_{132}\left( c,d\right) $ must vanish. The surviving terms for $Q_{222}\left(
a,b\right) :Q_{132}\left( c,d\right) $ are $(4/7)$ times the terms
\begin{equation}
\begin{array}{c}
-5\left( \widehat{{\mathbf a}}\cdot \widehat{{\mathbf d}}\right) ^{2}\left(
\widehat{{\mathbf c}}\cdot \widehat{{\mathbf d}}\right) +15\left(
\widehat{{\mathbf a}}\cdot \widehat{{\mathbf b}}\right) \left(
\widehat{{\mathbf c}}\cdot \widehat{{\mathbf d}}\right) \left(
\widehat{{\mathbf a}}\cdot \widehat{{\mathbf d}}\right) \left(
\widehat{{\mathbf b}}\cdot \widehat{{\mathbf d}}\right) \\
-5\left( \widehat{{\mathbf b}}\cdot \widehat{{\mathbf d}}\right) ^{2}\left(
\widehat{{\mathbf c}}\cdot \widehat{{\mathbf d}}\right) -3\left(
\widehat{{\mathbf a}}\cdot \widehat{{\mathbf b}}\right) ^{2}\left(
\widehat{{\mathbf c}}\cdot \widehat{{\mathbf
d}}\right) +2\left( \widehat{{\mathbf c}}\cdot \widehat{{\mathbf d}}\right) \\
+2\left( \widehat{{\mathbf a}}\cdot \widehat{{\mathbf c}}\right) \left(
\widehat{{\mathbf a}}\cdot \widehat{{\mathbf d}}\right) -3\left(
\widehat{{\mathbf a}}\cdot \widehat{{\mathbf b}}\right) \left(\widehat{{\mathbf
a}}\cdot \widehat{{\mathbf c}}\right) \left(
\widehat{{\mathbf b}}\cdot \widehat{{\mathbf d}}%
\right) \\ -3\left( \widehat{{\mathbf a}}\cdot \widehat{{\mathbf b}}\right)
\left( \widehat{{\mathbf a}}\cdot \widehat{{\mathbf d}}\right) \left(
\widehat{{\mathbf b}}\cdot \widehat{{\mathbf c}}\right) +2\left(
\widehat{{\mathbf b}}\cdot \widehat{{\mathbf c}}\right) \left(
\widehat{{\mathbf b}}\cdot \widehat{{\mathbf d}}\right) \ ,
\end{array}
\label{eq64}
\end{equation}
Inserting into Eq.\,(\ref{eq61}) gives the final answer for the fourfold coupling
shown in the Eq.\,(\ref{eqA9}).

Evidently the above procedure will work for spherical harmonics of
arbitrarily high rank and argument, coupled to each other any number of
times. The algorithm is susceptible to algebraic coding within a reasonably
sophisticated algebraic manipulation program.

In the Appendix of this paper, we give results for a selection of spherical
harmonic couplings as a reference and as a check of the implementation of
our method.

\section{CONCLUSIONS}

Although a large body of work covers angular coupling of irreducible
tensors, explicit results for the coupling of Cartesian tensors of arbitrary
rank have not been available. For many physical applications, using
Cartesian coupling has some distinct advantages over the corresponding
spherical case. We have shown that the Cartesian coupling of spherical
harmonics can be performed in a straightforward manner, following a well-defined
procedure. The results are relatively simple and easy to interpret.
Specifically, a simple algorithm permits one to write down directly a scalar
expression for the coupling to zero of any number of spherical harmonics in
terms of the unit vectors involved. We also note that leaving the
coupling to a numerical calculation of azimuthal sums can introduce significant
numerical errors when many intermediate terms should add to zero, but
do not because of numerical truncations. This
difficulty does not arise when resultant analytic forms are first
calculated, as in this paper, before numerics are programmed.

Note added in proof: After this manuscript was submitted, R. F. Snider brought
to our attention earlier work on irreducible Cartesian tensors that the reader
may find useful.\footnote{J. A. R. Coope, R. F. Snider, and F. R. McCourt,
{\bf J. Chem. Phys. 43}, 2269 (1965 ); J. A. R. Coope and R. F. Snider,
{\bf J. Math. Phys. 11}, 1003 (1970); J. A. R. Coope, {\it ibid.} 11, 1591 (1970).}

\section{Acknowledgment}

The work of the authors is supported in part by the U. S. Department of
Energy under grant No. DE-FG05-86ER40270.

\appendix

\section{Appendix}

In this appendix, we give examples of the reduction described in the paper
for some commonly found spherical harmonic couplings to a scalar. The
results serve to show the simplicity of the expressions, to exhibit their
usefulness for physical interpretations in terms of the initial vector
directions contained in the spherical harmonics, and to act as reference.

[Note: We have suppressed bold facing and vector hats in the following.
Non-the-less, the letters $a,b,c,d,\cdots$ should be taken as vectors
of unit length.]

\begin{equation} \begin{array}{l}
\left[ Y^{\left[ 2\right]} (a)\times \left[ Y^{\left[ 1\right]} (b)\times
Y^{\left[ 1\right]} (c)\right] ^{\left[ 2\right]} \right] ^{\left[ 0\right]} =
3/(16\pi ^{3/2})\\ \{ 5(a\cdot c)^2(b\cdot c)\\ -2(a\cdot b)(a\cdot c)\\
-(b\cdot c)\} \, . \label{eqA1}
\end{array} \end{equation}

\begin{equation} \begin{array}{l}
\left[ \left[ Y^{\left[ 2\right]} (a)\times Y^{\left[ 2\right]} (b)\right]
^{\left[ 2\right]} \times Y^{\left[ 2\right]} (c)\right] ^{\left[ 0\right]} =
5/(8\sqrt{14}\pi ^{3/2})\\ \{ 9(a\cdot b)(a\cdot c)(b\cdot c)\\ -3(a\cdot b)^2\\
-3(a\cdot c)^2\\ -3(b\cdot c)^2\\ +2\} \, . \label{eqA2}
\end{array} \end{equation}

\begin{equation} \begin{array}{l}
\left[ Y^{\left[ 1\right]} (a)\times \left[ Y^{\left[ 1\right]} (b)\times
Y^{\left[ 2\right]} (c)\right] ^{\left[ 1\right]} \right] ^{\left[ 0\right]} =
\sqrt{3}/(8\sqrt{2}\pi ^{3/2})\\ \{ 3(a\cdot c)(b\cdot c)\\ - (a\cdot b)\} \, .
\label{eqA3}
\end{array} \end{equation}

\begin{equation} \begin{array}{l}
\left[ \left[ Y^{\left[ 1\right]} (a)\times Y^{\left[ 1\right]} (b)\right]
^{\left[ 2\right]} \times \left[ Y^{\left[ 1\right]} (c)\times Y^{\left[
1\right]} (d)\right] ^{\left[ 2\right]} \right] ^{\left[ 0\right]} =
3/(8\sqrt{2}\pi^{2})\\ \{ 3(a\cdot c)(b\cdot d)\\ +3(b\cdot c)(a\cdot d)\\
-2(a\cdot b)(c\cdot d)\} \, . \label{eqA4}
\end{array} \end{equation}

\begin{equation}
\begin{array}{l}
\left[ \left[ Y^{\left[ 1\right]} (a)\times Y^{\left[ 1\right]} (b)\right]
^{\left[ 2\right]} \times \left[ Y^{\left[ 1\right]} (c)\times Y^{\left[
3\right]} (d)\right] ^{\left[ 2\right]} \right] ^{\left[ 0\right]}=
3\sqrt{15}/(80\sqrt{2}\pi ^{2}) \\
\{ 5(a\cdot c)(b\cdot c)(c\cdot d)\\ -(a\cdot c)(b\cdot d)\\ -(b\cdot c)(a\cdot
d)\\ -(c\cdot d)(a\cdot b)\} . \label{eqA5}
\end{array} \end{equation}

\begin{equation} \begin{array}{l}
\left[ \left[ Y^{\left[ 1\right]} (a)\times Y^{\left[ 3\right]} (b)\right]
^{\left[ 2\right]} \times \left[ Y^{\left[ 1\right]}(c)\times Y^{\left[
3\right]}(d)\right] ^{\left[ 2\right]} \right] ^{\left[ 0\right]}=
3\sqrt{50}/(160\pi ^{2}) \\
\{ -10(c\cdot d)(a\cdot d)(b\cdot d)\\ +25(c\cdot d)(b\cdot d)^2(a\cdot
b)\\ -3(c\cdot d)(a\cdot b) \\
+2(a\cdot c)(b\cdot d)\\ +2(b\cdot c)(a\cdot d)\\ -10(b\cdot c)(b\cdot
d)(a\cdot b)\} \, .  \label{eqA6}
\end{array} \end{equation}

\begin{equation} \begin{array}{l}
\left[ \left[ Y^{\left[ 2\right]} (a)\times Y^{\left[ 2\right]} (b)\right]
^{\left[ 1\right]} \times \left[ Y^{\left[ 1\right]}(c)\times Y^{\left[
1\right]} (d)\right] ^{\left[ 1\right]} \right] ^{\left[ 0\right]} =
3\sqrt{15}/(32\pi ^{2})\\ (a\cdot b)(a\cdot c) \\ \{ (b\cdot d)\\ -(b\cdot
c)(a\cdot d)\} \, . \label{eqA7}
\end{array} \end{equation}

\begin{equation} \begin{array}{l}
\left[ \left[ Y^{\left[ 2\right]} (a)Y^{\left[ 2\right]} (b)\right] ^{\left[
2\right]} \times \left[ Y^{\left[ 1\right]} (c)\times Y^{\left[ 1\right]}
(d)\right] ^{\left[ 2\right]} \right] ^{\left[ 0\right]} =
\sqrt{15}/(32\sqrt{7}\pi ^{2})\\ \{ -6(c\cdot d)(a\cdot b)^2\\ +4(c\cdot d)\\
-6(a\cdot c)(a\cdot d)\\ +9(a\cdot c)(b\cdot d)(a\cdot b)\\ +9(b\cdot c)(a\cdot
d)(a\cdot b)\\ -6(b\cdot c)(b\cdot d)\} \, . \label{eqA8}
\end{array} \end{equation}

\begin{equation} \begin{array}{l}
\left[ \left[ Y^{\left[ 2\right]} (a)\times Y^{\left[ 2\right]} (b)\right]
^{\left[ 2\right]} \times \left[ Y^{\left[ 1\right]} (c)\times Y^{\left[
3\right]} (d)\right] ^{\left[ 2\right]} \right] ^{\left[ 0\right]} =
3\sqrt{5}/(16\sqrt{14}\pi ^{2})\\ \{ -5(c\cdot d)(a\cdot d)^2\\ +15(c\cdot
d)(a\cdot
d)(b\cdot d)(a\cdot b)\\ -5(c\cdot d)(b\cdot d)\\ -3(c\cdot d)(a\cdot b)^2\\
+2(c\cdot d)\\ +2(a\cdot c)(a\cdot d)\\ -3(a\cdot c)(b\cdot d)(a\cdot b)\\
-3(b\cdot c)(a\cdot d)(a\cdot b)\\ +2(b\cdot c)(b\cdot d)\} \, . \label{eqA9}
\end{array} \end{equation}

\begin{equation} \begin{array}{l}
\left[ \left[ \left[ Y^{\left[ 2\right]} (a)\times Y^{\left[ 1\right]}
(b)\right] ^{\left[ 1\right]} \times Y^{\left[ 2\right]} (c)\right] ^{\left[
1\right]} \left[ \times Y^{\left[ 2\right]} (d)\times Y^{\left[ 2\right]}
(e)\right] ^{\left[ 1\right]} \right] ^{\left[ 0\right]} =
15\sqrt{3}/(64\sqrt{2}\pi ^{5/2})\\ (a\cdot b)(d\cdot e)\\ \{ 3(a\cdot
c)(b\cdot
d)(c\cdot e)\\ -3(a\cdot c)(b\cdot e)(c\cdot d)\\ -3(a\cdot d)(b\cdot c)(c\cdot e)\\
+2(a\cdot d)(b\cdot e)\\ +3(a\cdot e)(b\cdot c)(c\cdot d)\\ -2(a\cdot e)(b\cdot
d)\} \, . \label{eqA10}
\end{array} \end{equation}

\begin{equation} \begin{array}{l}
\left[ \left[ \left[ Y^{\left[ 2\right]} (a)\times Y^{\left[ 2\right]}
(b)\right] ^{\left[ 1\right]} \times Y^{\left[ 2\right]} (c)\right] ^{\left[
2\right]} \times \left[ Y^{\left[ 2\right]} (d)\times Y^{\left[ 2\right]}
(e)\right] ^{\left[ 2\right]} \right] ^{\left[ 0\right]} =
15\sqrt{15}/(64\sqrt{14}\pi ^{5/2})\\ (a\cdot b)\\ \{ -2(a\cdot c)(b\cdot
d)(c\cdot d)\\ +3(a\cdot c)(b\cdot d)(c\cdot e)(d\cdot e)\\ +3(a\cdot c)(b\cdot
e)(c\cdot d)(d\cdot e)\\ -2(a\cdot c)(b\cdot e)(c\cdot e)\\ +2(a\cdot d)(b\cdot
c)(c\cdot d)\\ -3(a\cdot d)(b\cdot c)(c\cdot e)(d\cdot e)\\ -3(a\cdot e)(b\cdot
c)(c\cdot d)(d\cdot e)\\ +2(a\cdot e)(b\cdot c)(c\cdot e)\} \, . \label{eqAll}
\end{array} \end{equation}

\begin{equation} \begin{array}{l}
\left[ \left[ \left[ Y^{\left[ 2\right]} (a)\times Y^{\left[ 2\right]}
(b)\right] ^{\left[ 2\right]} \times Y^{\left[ 2\right]} (c)\right] ^{\left[
1\right]} \times \left[ Y^{\left[ 2\right]} (d)\times Y^{\left[ 2\right]}
(e)\right] ^{\left[ 1\right]}\right] ^{\left[ 0\right]} =
15\sqrt{15}/(64\sqrt{14}\pi^{5/2})\\ (d\cdot e)\\ \{ 3(a\cdot b)(a\cdot
c)(b\cdot d)(c\cdot e)\\ -3(a\cdot b)(a\cdot c)(b\cdot e)(c\cdot d)\\ +3(a\cdot
b)(a\cdot d)(b\cdot c)(c\cdot e)\\ -3(a\cdot b)(a\cdot e)(b\cdot c)(c\cdot d)\\
-2(a\cdot c)(a\cdot d)(c\cdot e)\\ +2(a\cdot c)(a\cdot e)(c\cdot d)\\ -2(b\cdot
c)(b\cdot d)(c\cdot e)\\ +2(b\cdot c)(b\cdot e)(c\cdot d)\} \, . \label{eqA12}
\end{array} \end{equation}

\begin{equation} \begin{array}{l}
\left[ \left[ \left[ Y^{\left[ 2\right]} (a)\times Y^{\left[ 2\right]}
(b)\right] ^{\left[ 2\right]} \times Y^{\left[ 2\right]} (c)\right] ^{\left[
2\right]} \times \left[ Y^{\left[ 2\right]} (d)\times Y^{\left[ 2\right]}
(e)\right] ^{\left[ 2\right]} \right] ^{\left[ 0\right]} =
 25/(448\sqrt{14} \pi^ {5/2})\\
\{ 36(a\cdot b)^2(c\cdot d)^2 -108(a\cdot b)^2(c\cdot d)(c\cdot e)(d\cdot
e)+36(a\cdot b)^2(c\cdot e)^2 +72(a\cdot b)^2(d\cdot e)^2  \\
-48(a\cdot b)^2 -108(a\cdot b)(a\cdot c)(b\cdot c)(d\cdot e)^2 +72(a\cdot
b)(a\cdot c)(b\cdot c)\\ -54(a\cdot b)(a\cdot c)(b\cdot d)(c\cdot d)
+81(a\cdot b)(a\cdot c)(b\cdot d)(c\cdot e)(d\cdot e) \\
+81(a\cdot b)(a\cdot c)(b\cdot e)(c\cdot d)(d\cdot e)
-54(a\cdot b)(a\cdot c)(b\cdot e)(c\cdot e)\\
-54(a\cdot b)(a\cdot d)(b\cdot c)(c\cdot d)
+81(a\cdot b)(a\cdot d)(b\cdot c)(c\cdot e)(d\cdot e)\\
+36(a\cdot b)(a\cdot d)(b\cdot d)
-54(a\cdot b)(a\cdot d)(b\cdot e)(d\cdot e) \\
+81(a\cdot b)(a\cdot e)(b\cdot c)(c\cdot d)(d\cdot e)
-54(a\cdot b)(a\cdot e)(b\cdot c)(c\cdot e) \\
-54(a\cdot b)(a\cdot e)(b\cdot d)(d\cdot e)
+36(a\cdot b)(a\cdot e)(b\cdot e) \\
+36(a\cdot c)^2(d\cdot e)^2-24(a\cdot c)^2
+36(a\cdot c)(a\cdot d)(c\cdot d)\\
-54(a\cdot c)(a\cdot d)(c\cdot e)(d\cdot e)
-54(a\cdot c)(a\cdot e)(c\cdot d)(d\cdot e)\\
+36(a\cdot c)(a\cdot e)(c\cdot e)-12(a\cdot d)^2
+36(a\cdot d)(a\cdot e)(d\cdot e)-12(a\cdot e)^2\\
+36(b\cdot c)^2(d\cdot e)^2-24(b\cdot c)^2
+36(b\cdot c)(b\cdot d)(c\cdot d)\\
+54(b\cdot c)(b\cdot d)(c\cdot e)(d\cdot e)
-54(b\cdot c)(b\cdot e)(c\cdot d)(d\cdot e) \\
+36(b\cdot c)(b\cdot e)(c\cdot e)-12(b\cdot d)^2 +36(b\cdot d)(b\cdot e)(d\cdot
e)-12(b\cdot e)^2\\ -24(c\cdot d)^2 +72(c\cdot d)(c\cdot e)(d\cdot e)
-24(c\cdot e)^2-48(d\cdot e)^2 +32\} \, . \label{eqA13}
\end{array} \end{equation}

\begin{equation} \begin{array}{l}
\left[ \left[ \left[ Y^{\left[ 2\right]} (a)\times Y^{\left[ 2\right]}
(b)\right] ^{\left[ 1\right]} \times Y^{\left[ 2\right]} (c)\right] ^{\left[
1\right]} \times \left[ Y^{\left[ 1\right]} (d)\times Y^{\left[ 1\right]}
(e)\right] ^{\left[ 1\right]} \right] ^{\left[ 0\right]} =
3\sqrt{15}/(64\sqrt{2}\pi ^5/2)\\ (a\cdot b)\\ \{ 3(a\cdot c)(b\cdot d)(c\cdot e)\\
-3(a\cdot c)(b\cdot e)(c\cdot d)\\ -3(a\cdot d)(b\cdot c)(c\cdot e)\\ +2(a\cdot
d)(b\cdot e)\\ +3(a\cdot e)(b\cdot c)(c\cdot d)\\ -2(a\cdot e)(b\cdot d)\} \, .
\label{eqA14}
\end{array} \end{equation}

\begin{equation} \begin{array}{l}
\left[ \left[ \left[ Y^{\left[ 2\right]} (a)\times Y^{\left[ 2\right]}
(b)\right] ^{\left[ 1\right]} \times Y^{\left[ 2\right]} (c)\right] ^{\left[
2\right]} \times \left[ Y^{\left[ 1\right]} (d)\times Y^{\left[ 1\right]}
(e)\right] ^{\left[ 2\right]} \right] ^{\left[ 0\right]} =
9\sqrt{5}/(64\sqrt{2}\pi ^{5/2})\\ (a\cdot b)\\ \{ (a\cdot c)(b\cdot d)(c\cdot e)\\
+(a\cdot c)(b\cdot e)(c\cdot d)\\ -(a\cdot d)(b\cdot c)(c\cdot e)\\ -(a\cdot
e)(b\cdot c)(c\cdot d)\} \, . \label{eqA15}
\end{array} \end{equation}

\begin{equation} \begin{array}{l}
\left[ \left[ \left[ Y^{\left[ 2\right]} (a)\times Y^{\left[ 2\right]}
(b)\right] ^{\left[ 2\right]} \times Y^{\left[ 2\right]} (c)\right] ^{\left[
1\right]} \times \left[ Y^{\left[ 1\right]} (d)\times Y^{\left[ 1\right]}
(e)\right] ^{\left[ 1\right]} \right] ^{\left[ 0\right]} =
15\sqrt{3}/(64\sqrt{14}\pi ^{5/2})\\ \{ 3(a\cdot b)(a\cdot c)(b\cdot d)(c\cdot e)\\
-3(a\cdot b)(a\cdot c)(b\cdot e)(c\cdot d)\\ +3(a\cdot b)(a\cdot d)(b\cdot
c)(c\cdot e)\\ -3(a\cdot b)(a\cdot e)(b\cdot c)(c\cdot d)\\ -2(a\cdot c)(a\cdot
d)(c\cdot e)\\ +2(a\cdot c)(a\cdot e)(c\cdot d)\\ -2(b\cdot c)(b\cdot d)(c\cdot
e)\\ +2(b\cdot c)(b\cdot e)(c\cdot d)\} \, . \label{eqA16}
\end{array} \end{equation}

\begin{equation} \begin{array}{l}
\left[ \left[ \left[ Y^{\left[ 2\right]} (a)\times Y^{\left[ 2\right]}
(b)\right] ^{\left[ 2\right]} \times Y^{\left[ 2\right]} (c)\right] ^{\left[
2\right]} \times \left[ Y^{\left[ 1\right]} (d)\times Y^{\left[ 1\right]}
(e)\right] ^{\left[ 2\right]} \right] ^{\left[ 0\right]} =
5\sqrt{3}/(448\sqrt{2}\pi ^{5/2})\\ \{ -36(a\cdot b)^2(c\cdot d)(c\cdot e)\\
+24(a\cdot b)^2(d\cdot e)\\ -36(a\cdot b)(a\cdot c)(b\cdot c)(d\cdot e)\\
+27(a\cdot b)(a\cdot c)(b\cdot d)(c\cdot e)\\ +27(a\cdot b)(a\cdot c)(b\cdot
e)(c\cdot d)\\ +27(a\cdot b)(a\cdot d)(b\cdot c)(c\cdot e)\\ -18(a\cdot b)(a\cdot
d)(b\cdot e)\\ +27(a\cdot b)(a\cdot e)(b\cdot c)(c\cdot d)\\ -18(a\cdot
b)(a\cdot e)(b\cdot d)\\ +12(a\cdot c)^2(d\cdot e)\\ -18(a\cdot c)(a\cdot
d)(c\cdot e)\\ -18(a\cdot c)(a\cdot e)(c\cdot d)\\ +12(a\cdot d)(a\cdot e)\\
+12(b\cdot c)^2(d\cdot e)\\ -18(b\cdot c)(b\cdot d)(c\cdot e)\\ -18(b\cdot
c)(b\cdot e)(c\cdot d)\\ +12(b\cdot d)(b\cdot e)\\ +24(c\cdot d)(c\cdot e)\\
-16(d\cdot e)\} \, . \label{eqA17}
\end{array} \end{equation}

\begin{equation} \begin{array}{l}
\left[ \left[ \left[ Y^{\left[ 2\right]} (a)\times Y^{\left[ 2\right]}
(b)\right] ^{\left[ 1\right]} \times Y^{\left[ 2\right]} (c)\right] ^{\left[
2\right]} \times \left[ Y^{\left[ 1\right]} (d)\times Y^{\left[ 3\right]}
(e)\right] ^{\left[ 2\right] }\right] ^{\left[ 0\right]} =3\sqrt{15}/(64\pi
^{5/2})\\ (a\cdot b)\\ \{ -(a\cdot c)(b\cdot d)(c\cdot e)\\ -(a\cdot c)(b\cdot
e)(c\cdot d)\\ +5(a\cdot c)(b\cdot e)(c\cdot e)(d\cdot e)\\ +(a\cdot d)(b\cdot
c)(c\cdot e)\\ +(a\cdot e)(b\cdot c)(c\cdot d)\\ -5(a\cdot e)(b\cdot c)(c\cdot
e)(d\cdot e)\} \, . \label{eqA18}
\end{array} \end{equation}

\begin{equation} \begin{array}{l}
\left[ \left[ \left[ Y^{\left[ 2\right]} (a)\times Y^{\left[ 2\right]}
(b)\right] ^{\left[ 2\right]} \times Y^{\left[ 2\right]} (c)\right] ^{\left[
2\right]} \times \left[ Y^{\left[ 1\right]} (d)\times Y^{\left[ 3\right]}
(e)\right] ^{\left[ 2\right]} \right] ^{\left[ 0\right]} =15/(448\pi ^{5/2})\\
\{
12(a\cdot b)^2(c\cdot d)(c\cdot e)\\ -30(a\cdot b)^2(c\cdot e)^2(d\cdot e)\\
+12(a\cdot b)^2(d\cdot e)\\ -18(a\cdot b)(a\cdot c)(b\cdot c)(d\cdot e)\\
-9(a\cdot
b)(a\cdot c)(b\cdot d)(c\cdot e)\\ -9(a\cdot b)(a\cdot c)(b\cdot e)(c\cdot d)\\
+45(a\cdot b)(a\cdot c)(b\cdot e)(c\cdot e)(d\cdot e)\\ -9(a\cdot b)(a\cdot
d)(b\cdot c)(c\cdot e)\\ +6(a\cdot b)(a\cdot d)(b\cdot e)\\ -9(a\cdot b)(a\cdot
e)(b\cdot c)(c\cdot d)\\ +45(a\cdot b)(a\cdot e)(b\cdot c)(c\cdot e)(d\cdot
e)\\ +6(a\cdot b)(a\cdot e)(b\cdot d)\\ -30(a\cdot b)(a\cdot e)(b\cdot
e)(d\cdot e)\\ +6(a\cdot c)^2(d\cdot e)\\ +6(a\cdot c)(a\cdot d)(c\cdot e)\\
+6(a\cdot c)(a\cdot e)(c\cdot d)\\ -30(a\cdot c)(a\cdot e)(c\cdot e)(d\cdot
e)\\ -4(a\cdot d)(a\cdot e)\\ +10(a\cdot e)^2(d\cdot e)\\ +6(b\cdot c)^2(d\cdot e)\\
+6(b\cdot c)(b\cdot d)(c\cdot e)\\ +6(b\cdot c)(b\cdot e)(c\cdot d)\\
-30(b\cdot c)(b\cdot e)(c\cdot e)(d\cdot e)\\ -4(b\cdot d)(b\cdot e)\\
+10(b\cdot e)^2(d\cdot e)\\ -8(c\cdot d)(c\cdot e)\\ +20(c\cdot e)^2(d\cdot e)\\
-8(d\cdot e)\} \, . \label{eqA19}
\end{array} \end{equation}

\begin{equation} \begin{array}{l}
\left[ \left[ \left[ Y^{\left[ 1\right]} (a)\times Y^{\left[ 2\right]}
(b)\right] ^{\left[ 2\right]} \times Y^{\left[ 2\right]} (c)\right] ^{\left[
2\right]} \times \left[ Y^{\left[ 1\right]} (d)\times Y^{\left[
1\right]}(e)\right] ^{\left[ 2\right]} \right] ^{\left[ 0\right]} =
3\sqrt{3}/(64\sqrt{2}\pi ^{5/2})\\ \{ 3(a\cdot c)(b\cdot d)(c\cdot e)\\ -3(a\cdot
c)(b\cdot e)(c\cdot d)\\ -3(a\cdot d)(b\cdot c)(c\cdot e)\\ +2(a\cdot d)(b\cdot e)\\
+3(a\cdot e)(b\cdot c)(c\cdot d)\\ -2(a\cdot e)(b\cdot d)\} \, . \label{eqA20}
\end{array} \end{equation}

\begin{equation} \begin{array}{l}
\left[ \left[ \left[ Y^{\left[ 1\right]} (a)\times Y^{\left[ 1\right]}
(b)\right] ^{\left[ 2\right]} \times Y^{\left[ 2\right]} (c)\right] ^{\left[
2\right]} \times \left[ Y^{\left[ 1\right]} (d)\times Y^{\left[ 1\right]}
(e)\right] ^{\left[ 2\right]} \right] ^{\left[ 0\right]} =3/(64\sqrt{14}\pi
^{5/2})\\ \{ -12(a\cdot b)(c\cdot d)(c\cdot e)\\ +8(a\cdot b)(d\cdot e)\\
-12(a\cdot c)(b\cdot c)(d\cdot e)\\ +9(a\cdot c)(b\cdot d)(c\cdot e)\\
 +9(a\cdot c)(b\cdot e)(c\cdot d)\\ +9(a\cdot d)(b\cdot c)(c\cdot e)\\
-6(a\cdot d)(b\cdot e)\\ +9(a\cdot e)(b\cdot c)(c\cdot d)\\ -6(a\cdot e)(b\cdot
d)\} \, . \label{eqA21}
\end{array} \end{equation}

\begin{equation} \begin{array}{l}
\left[ \left[ \left[ Y^{\left[ 1\right]} (a)\times Y^{\left[ 1\right]}
(b)\right] ^{\left[ 1\right]} \times Y^{\left[ 2\right]} (e)\right] ^{\left[
2\right]} \times \left[ Y^{\left[ 1\right]} (d)\times Y^{\left[ 3\right]}
(e)^{\left[ 2\right]} \right] \right] ^{\left[ 0\right]} =3\sqrt{3}/(64\pi
^{5/2})\\ \{ -(a\cdot c)(b\cdot d)(c\cdot e)\\ -(a\cdot c)(b\cdot e)(c\cdot d)\\
+5(a\cdot e)(b\cdot e)(c\cdot e)(d\cdot e)\\ +(a\cdot d)(b\cdot c)(c\cdot e)\\
+(a\cdot e)(b\cdot c)(c\cdot d)\\ -5(a\cdot e)(b\cdot c)(c\cdot e)(d\cdot e)\}
\, . \label{eqA22}
\end{array} \end{equation}

\begin{equation} \begin{array}{l}
\left[ \left[ \left[ Y^{\left[ 1\right]} (a)\times Y^{\left[ 1\right]}
(b)\right] ^{\left[ 2\right]} \times Y^{\left[ 1\right]} (e)\right] ^{\left[
2\right]} \times \left[ Y^{\left[ 1\right]} (d)\times Y^{\left[ 3\right]}
(e)\right] ^{\left[ 2\right]} \right] ^{\left[ 0\right]} =
3\sqrt{3}/(64\sqrt{7}\pi ^{5/2})\\ \{ 4(a\cdot b)(c\cdot d)(c\cdot e)\\
-10(a\cdot b)(c\cdot e)^2(d\cdot e)\\ +4(a\cdot b)(d\cdot e)\\ -6(a\cdot
c)(b\cdot
c)(d\cdot e)\\ -3(a\cdot c)(b\cdot d)(c\cdot e)\\ -3(a\cdot c)(b\cdot e)(c\cdot d)\\
+15(a\cdot c)(b\cdot e)(c\cdot e)(d\cdot e)\\ -3(a\cdot d)(b\cdot c)(c\cdot e)\\
+2(a\cdot d)(b\cdot e)\\ -3(a\cdot e)(b\cdot c)(c\cdot d)\\ +15(a\cdot
e)(b\cdot c)(c\cdot e)(d\cdot e)\\ +2(a\cdot e)(b\cdot d)\\ -10(a\cdot
e)(b\cdot e)(d\cdot e)\} \, . \label{eqA23}
\end{array} \end{equation}

\begin{equation} \begin{array}{l}
\left[ \left[ \left[ Y\left[ 1\right] (a)\times Y^{\left[ 3\right]} (b)\right]
^{\left[ 2\right]} \times Y^{\left[ 2\right]} (e)\right] ^{\left[ 1\right]}
\times \left[ Y^{\left[ 1\right]} (d)\times Y^{\left[ 1\right]} (e)\right]
^{\left[ 1\right]} \right] ^{\left[ 0\right]} =3\sqrt{3}/(64\pi ^{5/2})\\ \{
5(a\cdot b)(b\cdot c)(b\cdot d)(c\cdot e)\\ -5(a\cdot b)(b\cdot c)(b\cdot e)(c\cdot d)\\
-(a\cdot c)(b\cdot d)(c\cdot e)\\ +(a\cdot c)(b\cdot e)(c\cdot d)\\ -(a\cdot
d)(b\cdot c)(c\cdot e)\\ +(a\cdot e)(b\cdot c)(c\cdot d)\} \, . \label{eqA24}
\end{array} \end{equation}

\begin{equation} \begin{array}{l}
\left[ \left[ \left[ Y^{\left[ 1\right]} (a)\times Y^{\left[ 3\right]}
(b)\right] ^{\left[ 2\right]} \times Y^{\left[ 2\right]} (c)\right] ^{\left[
2\right]} \times \left[ Y^{\left[ 1\right]} (d)\times Y^{\left[ 1\right]}
(e)\right] ^{\left[ 2\right]} \right] ^{\left[ 0\right]}
=
3\sqrt{3}/(64\sqrt{7}\pi ^{5/2})\\ \{ -10(a\cdot b)(b\cdot c)^2(d\cdot e)\\
+15(a\cdot b)(b\cdot c)(b\cdot d)(c\cdot e)\\ +15(a\cdot b)(b\cdot c)(b\cdot
e)(c\cdot d)\\ -10(a\cdot b)(b\cdot d)(b\cdot e)\\ -6(a\cdot b)(c\cdot d)(c\cdot e)\\
+4(a\cdot b)(d\cdot e)\\ +4(a\cdot c)(b\cdot c)(d\cdot e)\\ -3(a\cdot c)(b\cdot
d)(c\cdot e)\\ -3(a\cdot c)(b\cdot e)(c\cdot d)\\ -3(a\cdot d)(b\cdot c)(c\cdot e)\\
+2(a\cdot d)(b\cdot e)\\ -3(a\cdot e)(b\cdot c)(c\cdot d)\\ +2(a\cdot e)(b\cdot
d)\} \, . \label{eqA25}
\end{array} \end{equation}

\begin{equation} \begin{array}{l}
\left[ \left[ \left[ Y^{\left[ 1\right]} (a)\times Y^{\left[ 3\right]}
(b)\right] ^{\left[ 2\right]} \times Y^{\left[ 2\right]} (c)\right] ^{\left[
2\right]} \times \left[ Y^{\left[ 1\right]} (d)\times Y^{\left[ 3\right]}
(e)\right] ^{\left[ 2\right]} \right] ^{\left[ 0\right]}
=3/(32\sqrt{14}\pi
^{5/2})\\ \{ -15(a\cdot b)(b\cdot e)^2(d\cdot e)\\ -15(a\cdot b)(b\cdot
c)(b\cdot d)(c\cdot e)\\ -15(a\cdot b)(b\cdot c)(b\cdot e)(c\cdot d)\\
+75(a\cdot b)(b\cdot c)(b\cdot e)(c\cdot e)(d\cdot e)\\ +10(a\cdot b)(b\cdot
d)(b\cdot e)\\ -25(a\cdot b)(b\cdot e)^2(d\cdot e)\\ +6(a\cdot
b)(c\cdot d)(c\cdot e)\\ -15(a\cdot b)(c\cdot e)^2(d\cdot e)\\ +6(a\cdot b)(d\cdot e)\\
+6(a\cdot c)(b\cdot c)(d\cdot e)\\ +3(a\cdot c)(b\cdot d)(c\cdot e)\\ +3(a\cdot
c)(b\cdot e)(c\cdot d)\\ -15(a\cdot c)(b\cdot e)(c\cdot e)(d\cdot e)\\ +3(a\cdot
d)(b\cdot c)(c\cdot e)\\ -2(a\cdot d)(b\cdot e)\\ +3(a\cdot e)(b\cdot c)(c\cdot
d)\\ -15(a\cdot e)(b\cdot c)(c\cdot e)(d\cdot e)\\ -2(a\cdot e)(b\cdot d)\\
+10(a\cdot e)(b\cdot e)(d\cdot e)\} \, . \label{eqA26}
\end{array} \end{equation}

\bigskip

\end{document}